\documentclass[aps,twocolumn,showpacs,showkeys,superscriptaddress,prl,reprint, letterpaper]{revtex4-1}
\usepackage{graphicx}
\usepackage{dcolumn}
\usepackage{bm}
\usepackage{color}
\usepackage{ulem}

\begin{document}

\author{G. Bertaina}
\affiliation{INO-CNR BEC Center and Dipartimento di Fisica, Universit\`a di Trento, 38123 Povo, Italy}
\affiliation{Institute of Theoretical Physics, Ecole Polytechnique F\'{e}d\'{e}rale de Lausanne EPFL, CH-1015 Lausanne, Switzerland \\ ()}

\author{S. Giorgini}
\affiliation{INO-CNR BEC Center and Dipartimento di Fisica, Universit\`a di Trento, 38123 Povo, Italy}

\title{BCS-BEC crossover in a two-dimensional Fermi gas} 

\begin{abstract} 
We investigate the crossover from Bardeen-Cooper-Schrieffer (BCS) superfluidity to Bose-Einstein condensation (BEC) in a two-dimensional Fermi gas at $T=0$ using the fixed-node diffusion Monte Carlo method. We calculate the equation of state and the gap parameter as a function of the interaction strength, observing large deviations compared to mean-field predictions. In the BEC regime our results show the important role of dimer-dimer and atom-dimer interaction effects that are completely neglected in the mean-field picture. Results on Tan's contact parameter associated with short-range physics are also reported along the BCS-BEC crossover.    
\end{abstract}

\maketitle 

The study of ultracold atomic Fermi gases has become an active and rich field of research~\cite{RMP}. Important areas of investigation include the BCS-BEC crossover in a superfluid gas with resonantly enhanced interactions, the Chandrasekhar-Clogston instability of the superfluid state when spin polarization is increased, the possible onset of itinerant ferromagnetism in a gas with repulsive interactions~\cite{Jo09} and the realization of the Hubbard model for fermions loaded in optical lattices~\cite{RMP-Bloch}.  

Low dimensional configurations of degenerate Fermi gases have also been the object of experimental and theoretical studies~\cite{RMP,RMP-Bloch}. In particular, a two-dimensional (2D) ultracold Fermi gas has been recently realized using a highly anisotropic pancake-shaped potential, and the density profile of the cloud has been measured using {\it in situ} imaging~\cite{Turlapov}. On the theoretical side, the evolution from a superfluid state with large Cooper pairs to one with tight molecules in a 2D system of attractive fermions was first investigated by Miyake~\cite{Miyake83} and later by Randeria and coworkers~\cite{Randeria} aiming to describe high-$T_c$ superconductors.  More recent studies address the problem of the superfluid transition~\cite{Petrov03,Zhang08}, of harmonic trapping~\cite{2Dtrap} and of population and mass imbalance~\cite{2Dimbalance}. These studies are in general based on perturbative or mean-field (MF) approaches  that are suitable in the regime of weak coupling, but are bound to break down for stronger interactions. 

In this Letter we provide the first determination using quantum Monte Carlo methods of the equation of state at $T=0$ of a homogeneous 2D Fermi gas in the BCS-BEC crossover.  We also obtain results for the pairing gap and the contact parameter as a function of the interaction strength. In the strong-coupling regime the emergence of interaction effects involving dimers produce large deviations compared to MF predictions. A similar study carried out in 3D~\cite{Astra04} has provided an important benchmark against which experimental determination of the equation of state, using measurements of the dispersion of collective modes~\cite{Altmeyer07} or of {\it in situ} density profiles~\cite{Navon10}, have been successfully compared.  Hopefully, our results will stimulate more experimental efforts towards the realization of a 2D Fermi gas in the strong-coupling regime by means, for example, of a Feshbach resonance to increase the interaction parameter~\cite{Turlapov}.

\begin{figure}[pt]
\begin{center}
\includegraphics[width=8cm]{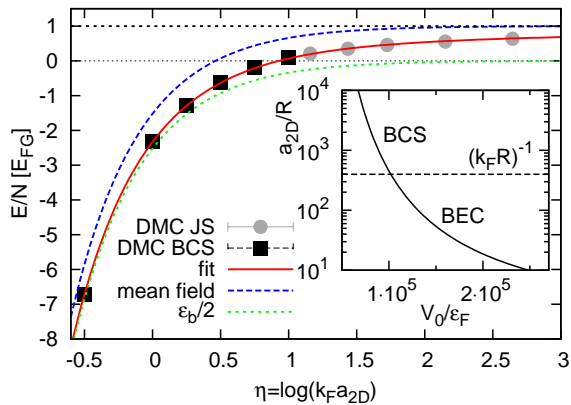}
\caption{\small{(color online).} Equation of state in the BCS-BEC crossover. Squares refer to the BCS and circles to the JS wave function.  The solid (red) line is a fit to the data, the dotted (green) line is half of the molecular binding energy and the dashed (blue) line is the MF prediction. The horizontal dotted (black) line denotes the energy per particle $E_{\text{FG}}$ of the noninteracting gas. Inset: 2D scattering length $a_{\text{2D}}$ as a function of the depth $V_0$ for a SW potential of radius $R$. The BCS and BEC regimes correspond, respectively, to $k_Fa_{\text{2D}} \gg 1$ and $k_Fa_{\text{2D}} \ll 1$.}
\label{fig1}
\end{center}
\end{figure}

We consider a homogeneous two-component Fermi gas described by the Hamiltonian
\begin{equation}
H=-\frac{\hbar^2}{2m}\left( \sum_{i=1}^{N_\uparrow}\nabla^2_i + \sum_{i^\prime=1}^{N_\downarrow}\nabla^2_{i^\prime}\right)
+\sum_{i,i^\prime}V(r_{ii^\prime}) \;,
\label{hamiltonian}
\end{equation}   
where $m$ denotes the mass of the particles, $i,j,...$ and $i^\prime,j^\prime,...$ label, respectively, spin-up and spin-down
particles and $N_\uparrow=N_\downarrow=N/2$, $N$ being the total number of atoms. We model the interspecies interatomic 
interactions using an attractive square-well (SW) potential: $V(r)=-V_0$ for $r<R$ ($V_0>0$), and $V(r)=0$ otherwise. 
In order to ensure that the mean interparticle distance is much larger than the range of the potential we use $nR^2=10^{-6}$, where $n$ is the gas number density, or equivalently $k_FR=0.0025$ in terms of the Fermi wave vector $k_F=\sqrt{2\pi n}$. 
We simulate a strictly 2D system and describe the low-energy collisions of the SW potential in terms of the 2D scattering length $a_{\text{2D}}$ defined as $a_{\text{2D}}=R\;e^{J_0(\kappa)/\kappa J_1(\kappa)}$,  
where $J_{0(1)}(x)$ are Bessel functions of the first kind and $\kappa=\sqrt{V_0mR^2/\hbar^2}$~\cite{note1}. The scattering length is non negative and diverges at $\kappa=0$ and at the zeros of $J_1$, corresponding to the appearance of new two-body bound states in the well. Close to these points the shallow dimers have size $a_{\text{2D}}$ and their binding energy is given by $\varepsilon_b=-4\hbar^2/(ma_{\text{2D}}^2e^{2\gamma})$, where $\gamma\simeq0.577$ is Euler's constant~\cite{note2}. The dependence of $a_{\text{2D}}$ on the depth $V_0$ in the region where the well supports only one bound state is shown in the inset of Fig.~\ref{fig1}. Two regions are clearly identified by comparing $a_{\text{2D}}$ with the mean interparticle distance $1/k_F$: i) $k_Fa_{\text{2D}}\gg 1$ corresponds to the BCS regime where interactions are weak and dimers are large and weakly bound, ii) $k_Fa_{\text{2D}}\ll 1$ corresponds to the BEC regime of tightly bound composite bosons. Compared to the 3D case the BCS-BEC crossover in 2D exhibits important differences. a) For a purely attractive potential a two-body bound state exists for arbitrarily weak attractions. b) The weak-coupling limit corresponds to a diverging scattering length $a_{\text{2D}}$. c) The 2D scattering amplitude of particles colliding at low energy is given by $f(k)=2\pi/[\log(2/ka_{\text{2D}}e^\gamma)+i\pi/2]$~\cite{Petrov01}. There is no range of values of $a_{\text{2D}}$ for which $f(k)$ is independent of interaction (unitary limit). d) The mean-field coupling constant can be written as $g=(2\pi\hbar^2/m)/\log(1/k_Fa_{\text{2D}})$  with logarithmic accuracy. Within the same accuracy, the region $k_Fa_{\text{2D}}\sim1$ identifies the strong-coupling crossover between the BCS and the BEC regimes [see inset of Fig.~\ref{fig1}].

Simulations are carried out in a square box of area $L^2=N/n$ with periodic boundary conditions, using the fixed-node diffusion Monte Carlo (FN-DMC) method. This numerical technique yields an upper bound for the ground-state energy of the gas, resulting from an ansatz for the nodal surface of the many-body wave function that is kept fixed during the calculation (see Ref.~\cite{Reynolds82} for more details). The boundary condition is enforced using a trial function that we choose of the general form~\cite{QMC} $\psi_T({\bf R})=\Phi_S({\bf R})\Phi_A({\bf R})$. $\Phi_S$ is a positive function of the particle coordinates ${\bf R}=({\bf r}_1,...,{\bf r}_N)$ and is symmetric in the exchange of particles with equal spin, while $\Phi_A$ satisfies the fermionic antisymmetry condition and determines the nodal surface of $\psi_T$.  The symmetric part is chosen of the Jastrow form $\Phi_S({\bf R})=\prod_{i,i^\prime}f_{\uparrow\downarrow}(r_{i i^\prime})$, where two-body correlation functions of the interparticle distance have been introduced for antiparallel spins. The $\Phi_A$ component is chosen as an antisymmetrized product  $\Phi_A({\bf R})={\cal A} \left( \phi(r_{11^\prime})\phi(r_{22^\prime})...\phi(r_{N_\uparrow N_\downarrow})\right)$ of pairwise orbitals of the form $\phi(r)=\beta\sum_{k_\alpha\leq k_F}e^{i{\bf k}_\alpha\cdot{\bf r}}+\varphi_s(r)$.  Here,  ${\bf k}_\alpha=2\pi/L(\ell_{\alpha x}\hat{x}+\ell_{\alpha y}\hat{y})$ indicate the plane-wave states in the box, with integers $\ell$'s summed up to the maximum value of the $k$-th shell accommodating $N/2$ particles, and $\beta$ is a variational parameter controlling the relative weight of the plane-wave sum to the spherical symmetric component $\varphi_s(r)$
. Two important regimes are described by the above trial wave function: i) if $\beta=0$ and $\varphi_s(r)=f_b(r)$ is the two-body bound state of the potential $V(r)$, $\psi_T({\bf R})$ describes a BCS state of dimers that is expected to be appropriate in the deep BEC regime; ii) if instead $\varphi_s=0$, the antisymmetric component in the trial function coincides with the product of the plane-wave Slater determinants for spin-up and spin-down particles, $\Phi_A({\bf R})=D_\uparrow(N_\uparrow)D_\downarrow(N_\downarrow)$~\cite{Bouchaud88}, and $\psi_T$ is a typical Jastrow-Slater (JS) function of a normal Fermi liquid. This description is expected to hold in the BCS regime of a weakly interacting gas where the effect of pairing on the ground-state energy is negligible. The more general form of the trial wave function written above interpolates between these two regimes. 

\begin{figure}[pb]
\begin{center}
\includegraphics[width=8cm]{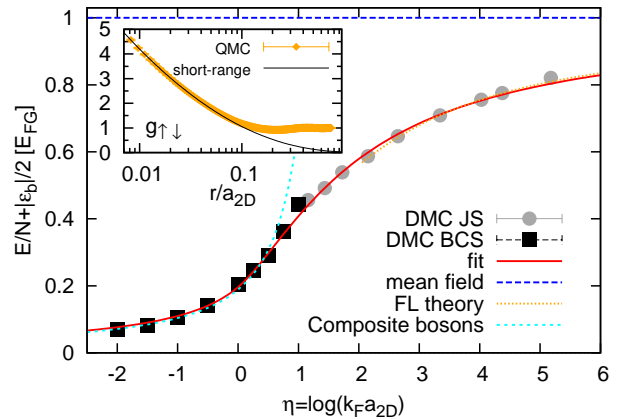}
\caption{\small{(color online).} Equation of state in the BCS-BEC crossover with $\varepsilon_b/2$ subtracted from $E/N$. Symbols are as in Fig.\ref{fig1}. The solid (red) line is a fit to the data, the other dotted lines show the equation of state (\ref{eosBose}) of composite bosons and the perturbation expansion holding in the BCS regime (see text). The dashed (blue) line is the MF result. Inset: short-range behavior of $g_{\uparrow\downarrow}$ for $\eta=2.15$.}
\label{fig2}
\end{center}
\end{figure}

\begin{table}[pt]
\caption{Energy per particle and molecular binding energy in the BEC-BCS crossover (energies are in units of $E_{\text{FG}}$).}
\begin{tabular}{cccc}
$\log(k_Fa_{\text{2D}})$ & $E/N$ & $\varepsilon_b/2$ & $E/N-\varepsilon_b/2$ \\ 
\colrule
-2.00  & -137.761(7)   & -137.832  &  0.070(7)   \\
-1.50  &   -50.593(4)   & -50.675     &  0.082(4)   \\
-1.00  &   -18.532(4)   & -18.637     &  0.105(4)   \\
-0.50  &     -6.714(4)   &  -6.856       &  0.142(4)   \\
 0.00  &     -2.318(2)   &   -2.522      &  0.204(2)    \\
 0.25  &     -1.283(12) &   -1.530      &  0.247(12) \\
 0.50  &     -0.638(10)  &  -0.928      &  0.290(10)    \\
 0.75  &     -0.201(12) &   -0.563      &  0.361(12)  \\
 1.44  &      0.349(6)    &   -0.143      &  0.492(6)    \\
 1.72  &      0.459(16)  &   -0.080      &  0.539(16)  \\
 2.15  &      0.552(2)    &   -0.034      &  0.587(2)     \\
 2.64  &      0.634(4)    &   -0.013      &  0.647(4)     \\
 3.34  &      0.706(2)    &   -0.003      &  0.709(2)     \\
 4.03  &      0.755(4)    &    0.000      &  0.755(4)     \\
 4.37  &      0.775(1)    &    0.000      &  0.775(1)     \\
 5.18  &      0.821(7)    &    0.000      &  0.821(7)     \\
\label{tab1}
\end{tabular}
\end{table}  
 
In Figs.~\ref{fig1}-\ref{fig2} and in Table~\ref{tab1} we report the FN-DMC results for the equation of state as a function of the interaction parameter in units of the energy per particle of the noninteracting gas $E_{\text{FG}}=\hbar^2k_F^2/4m=\varepsilon_F/2$,
where $\varepsilon_F$ is the Fermi energy. Calculations are carried out using $\psi_T$ of the BCS and JS form as described above
. The BCS wave function (corresponding to $\beta=0$) provides a lower energy for values of the interaction parameter $\eta=\log(k_Fa_{\text{2D}})\lesssim1$, while the JS
function (corresponding to $\beta\gg1$) is more favorable for larger values of $\eta$. The optimal parameter $\beta$ in the BCS orbital has been found to be zero even in the region $\eta\sim1$; finite values of $\beta$ have not given a significant improvement of the ground-state energy. The role of finite-size effects has been investigated by carrying out calculations with $N=26$ and  $N=98$. No significant change is seen when using the BCS trial function. In the case of the JS function a large suppression of such effects is obtained by using the theory of Fermi liquids. The difference in the energy per particle between the finite-size and the infinite system, in the interacting case, is assumed to be the same as in the noninteracting case, to lowest order in the effective mass (see~\cite{Lin01} for details). The MF result $E/N=E_{\text{FG}}+\varepsilon_b/2$~\cite{Miyake83,Randeria} is shown in Figs.~\ref{fig1}-\ref{fig2} for comparison. The inadequacy of the MF approach is best shown in Fig.~\ref{fig2}, where the molecular contribution is subtracted from the energy per particle. This figure has to be compared to Fig.~5 of Ref.~\cite{RMP}, concerning the 3D case: effects beyond mean-field  are much more pronounced in 2D than in 3D. In the BEC regime the FN-DMC results are fitted with the equation of state of a gas of composite bosons corresponding to hard disks of diameter $a_d$
\begin{eqnarray}
\frac{E}{N_d}+|\varepsilon_b|&=&\frac{2\pi\hbar^2n_d}{m_d}\frac{1}{\log(1/n_da_d^2)} \biggl[ 1-\frac{\log\log(1/n_da_d^2)}{\log(1/n_da_d^2)}
 \nonumber \\
&+& \frac{\log\pi+2\gamma+1/2}{\log(1/n_da_d^2)} \biggr] \;,
\label{eosBose}
\end{eqnarray}
where $m_d=2m$ is the mass of the dimer, while the number of dimers, and correspondingly their density $n_d$, is half of the total number of fermions $N_d=N/2$.  The above expression includes beyond mean-field terms~\cite{Bose2D} and allows for a precise determination of the dimer-dimer scattering length $a_d$. We obtain $a_d=0.55(4)a_{2D}$, in agreement with the four-body calculation in Ref.~\cite{Petrov03}. In the opposite BCS regime, where the contribution of the pairing gap can be neglected, the fermionic equation of state can be described in terms of an attractive normal Fermi liquid (FL). Beyond logarithmic accuracy one has the second-order expansion in terms of $\eta$~\cite{Bloom75,Engelbrecht92} $\frac{E}{N}=E_{\text{FG}}\left( 1-\frac{1}{\eta}+\frac{A}{\eta^2}\right)$.
From a best fit we find the result $A=0.06(2)$ for the coefficient of the second-order term~\cite{note5}.

\begin{figure}[pb]
\begin{center}
\includegraphics[width=8cm]{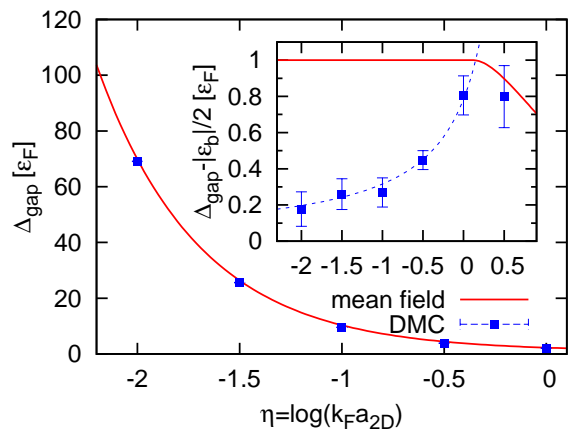}
\caption{\small{(color online).} Excitation gap in the BCS-BEC crossover. The solid (red) line is the MF result. Inset: excitation gap with $|\varepsilon_b|/2$ subtracted from $\Delta_{\text{gap}}$. The dashed (blue) line is a fit using the energy functional of a Fermi-Bose mixture.}
\label{fig3}
\end{center}
\end{figure}

In Fig.~\ref{fig3} we show the results for the pairing gap $\Delta_{\text{gap}}$ in the strong-coupling regime. This quantity is defined from the difference of ground-state energy $E(N_\uparrow,N_\downarrow)$ of systems having one and two more (less) particles $\Delta_{\text{gap}}=1/2[2E(N/2\pm1,N/2)-E(N/2\pm1,N/2\pm1)-E(N/2,N/2)]$ \cite{Carlson2003}. 
At the MF level~\cite{Miyake83,Randeria} the pairing gap coincides with the result for the order parameter $\Delta_{\text{gap}}=\Delta=\sqrt{2\varepsilon_F|\varepsilon_b|}$ if $|\varepsilon_b|<2\varepsilon_F$, and is given by $\Delta_{\text{gap}}=\varepsilon_F+|\varepsilon_b|/2$ for larger values of $|\varepsilon_b|$.  In the BEC regime the quantity $\Delta_{\text{gap}}-|\varepsilon_b|/2$, shown in the inset of Fig.~\ref{fig3}, displays the repulsive interaction effects between unpaired fermionic atoms and bosonic dimers. In fact, the energy of the system with one extra spin-up particle can be written as the sum of the contribution (\ref{eosBose}) of $N/2$ dimers and the Fermi-Bose interaction energy $E(N/2+1,N/2)=E(N/2,N/2)+g_{BF}n_d$, where $g_{BF}=3\pi\hbar^2/[m\log(1/n_d a_{ad}^2)]$ is the coupling constant fixed by the atom-dimer reduced mass $2m/3$ and the effective scattering length $a_{ad}$. By using the definition of $\Delta_{\text{gap}}$ and the value $a_d=0.55a_{\text{2D}}$ for the dimer-dimer scattering length in the energy functional (\ref{eosBose}), we find $a_{ad}=1.7(1)a_{\text{2D}}$ from the fit shown in the inset of Fig.~\ref{fig3}.  

\begin{figure}[pt]
\begin{center}
\includegraphics[width=8cm]{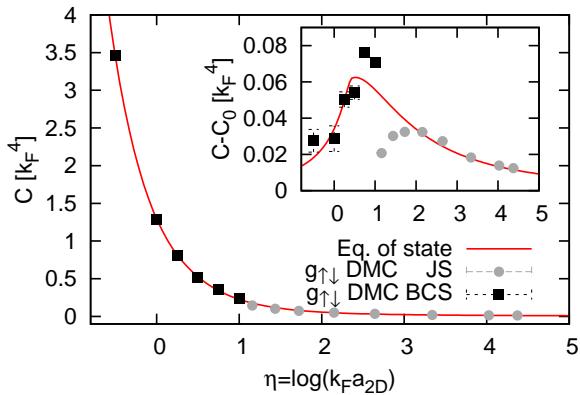}
\caption{Contact parameter in the BCS-BEC crossover. The solid line corresponds to the calculation from the derivative of the equation of state reported in Fig.\ref{fig1}. Inset: Contact parameter with the two-body contribution $C_0$ subtracted.}
\label{fig4}
\end{center}
\end{figure}

Finally, we calculate the contact parameter $C$~\cite{Tan08,Werner10,note6} defined from the short-range behavior of the antiparallel pair distribution function   $\lim_{r\to0}g_{\uparrow\downarrow}(r)=4C/k_F^4\log^2(r/a_{2D})$~[see inset of Fig.~\ref{fig2}]. The contact parameter is also related to the derivative of the equation of state with respect to the interaction parameter $C=(2\pi m/\hbar^2)d(nE/N)/d(\log k_Fa_{2D})$~\cite{Werner10}. The results are shown in Fig.~\ref{fig4}. In the inset we show the quantity $C-C_0$, where $C_0=(\pi m/\hbar^2)d(n\varepsilon_b)/d(\log k_Fa_{\text{2D}})$ is the contribution to the contact $C$ from the molecular state. The comparison between the two determinations of $C$ is a stringent consistency check of the theoretical approach. We find a good agreement with Tan's relation, apart from the region $\eta\sim1$ where small deviations are visible, both with the JS and BCS-type wave function, showing the need of a better optimization of $\psi_T$.

An important question relates to the relevance of these results for systems in harmonic traps. Two-dimensional configurations are realized if the transverse confinement is strong enough to reduce the kinematics to the $xy$-plane: $\hbar\omega_z\gg\varepsilon_F=\hbar\omega_\perp\sqrt{N}$, where we assumed isotropic trapping in the radial direction $\omega_x=\omega_y=\omega_\perp$. In these conditions the effective 2D scattering length can be expressed in terms of $a_{3D}$ and the transverse harmonic oscillator length $a_z=\sqrt{\hbar/m\omega_z}$ being given by $a_{\text{2D}}=a_z(2\sqrt{\pi/B}/e^\gamma)\exp(-\sqrt{\pi/2}a_z/a_{\text{3D}})$, where $B\simeq0.905$~\cite{Petrov01,RMP-Bloch}. For small, negative values of the 3D scattering length $a_{\text{3D}}$ the system is found in the BCS regime corresponding to an exponentially large $a_{\text{2D}}$. The BEC regime is reached if the absolute value of $a_{\text{3D}}$ is increased such that $|a_{\text{3D}}|\gg a_z/\log(1/k_Fa_z)$. An additional requirement concerns the dimer state, which is well described by the 2D expression only if $|\varepsilon_b|\ll\hbar\omega_z$~\cite{Petrov01}, or equivalently $a_{\text{2D}}\gg a_z$. We believe that this latter condition can be relaxed if, in the comparison with the results reported in this work, one considers quantities where the molecular contribution has been subtracted out.   

This work, as part of the European Science Foundation EUROCORES Program ``EuroQUAM-FerMix'', was supported by funds from the CNR and the EC Sixth Framework Programme.


\begin{thebibliography}{99}
\bibitem{note2} A different definition of $a_{\text{2D}}$ includes the factor $e^\gamma/2$, such that $\varepsilon_b=-\hbar^2/ma_{\text{2D}}^2$.
\bibitem{note5} Our value of $A$ is compatible with the result $A=3/4-\log2$ for the coefficient of the second-order term reported in Ref.~\cite{Engelbrecht92}. However, we disagree with Ref.~\cite{Engelbrecht92} on the value of the constant in the definition of $\eta$.
\bibitem{note6} The limit $r\to0$ is intended here as $R<r\ll k_F^{-1}$.
\end{thebibliography}
\end{document}